# Bio-Inspired Photonic Spectral Encoders

Yujia Zhang,[1,†] Xiangfu Lei,[2,†] Yinpeng Chen,[3,†] Chaojun Xu, [1,†] Hanxiao Cui,[2*] Tawfique Hasan, [4*] Yikai Su,[1] Zongyin Yang,[3*] Zhipei Sun,[5*] and Xuhan Guo[1*]

[1.] State Key Laboratory of Photonics and Communications, School of Information and Electronic Engineering, Shanghai Jiao Tong University; Shanghai, 200240, China.
[2.] School of Aeronautics and Astronautics, Sichuan University, Chengdu, 610065, China.
[3.] College of Information Science and Electronic Engineering, Zhejiang University, Hangzhou, 310027, China.
[4.] Department of Engineering, University of Cambridge; Cambridge, CB3 0FA, UK
[5.] Department of Electronics and Nanoengineering, Aalto University, Espoo 02150, Finland.
Corresponding authors: Hanxiao Cui, Tawfique Hasan, Zongying Yang, Zhipei Sun and Xuhan Guo
[†]These authors contributed equally to this work.

**Abstract**

Compact spectrometers promise to revolutionize sensing applications, offering a unique pathway to laboratory-grade analysis within a miniaturized footprint. Central to their performance is the encoding strategy to unknown spectra, which determines the efficiency, accuracy, and adaptability of spectral reconstruction. However, the absence of a unified spectral encoding framework has hindered the realization of optimal, high-performance compact spectrometers. We propose a transformative approach: an information-theoretic framework grounded in bio-inspired Bayesian expected information gain that defines the first generic light encoder for computational spectrometers. By optimizing three fundamental attributes at the lowest level of physical hierarchy, (1) orthogonality, (2) completeness, and (3) sparsity, we establish a design paradigm that transcends conventional encoding hardware limitations. We validate this paradigm with the first generic encoder capable of dynamically reconfiguring its response matrices. Experiments show superior reconstruction fidelity across diverse spectral regimes, enabling tunable spectral encoding tailored to varied input features. An ultra-high resolution of 6 pm and a broad measurable bandwidth of 30 nm are experimentally validated. By bridging the gap between theoretical encoding principles and reconfigurable hardware, our framework defines a coherent basis for future advances in compact spectrometry.

**Introduction**

Biological encoding systems excel at processing complex information with unparalleled energy efficiency and adaptability.[1] The sophistication of natural sensory organs and their underlying biological encoding mechanisms represents a pinnacle through millennia of evolutionary refinement. These systems, honed through millennia of natural selection, balance energy efficiency with the need for effective environmental interaction. Their performance is governed by three key principles: orthogonality, completeness, and sparsity. Orthogonality allows distinct sensory channels to operate with minimal overlap, as seen in marine mammals that communicate through specific, noninterfering frequencies.[2] Completeness enables organisms to integrate multiple sensory modalities for a comprehensive understanding of their environment[3], while sparsity ensures energy-efficient information processing by activating only essential pathways. The nematode Caenorhabditis elegans, for instance, performs complex behaviors such as signal compression and noise reduction with an exceptionally small number of neurons.[4] In contrast, artificial encoding systems, though capable of high task-specific performance, remain limited by rigid architectures that lack such versatility. Neural networks partially emulate biological encoding through backpropagation, yet their reliance on large datasets and high power consumption exposes a critical efficiency gap.[5,6] Bridging this gap requires reimagining encoding at the hardware level, where biological efficiency can be translated into scalable and adaptive engineering.

In previous computational spectrometer researches, majority of them focus on enhancing key performance metrics including footprints, resolution, and bandwidth by diverse material platforms[7-14] or device architectures[15-24]. On the other hand, recently in 2025, self-adaptive algorithms are emerging to improve the accuracy of reconstruction, while these algorithms are currently only employed in the post-processing step of the software[25,26]. Unlike prior task-specific or software-dependent approaches[27], we propose a photonic encoder systems that achieves hardware-level adaptability by integrating three bio-inspired design axioms, orthogonality, completeness, and sparsity, into reconfigurable optical components. The encoder consists of a high-Q Mach-Zehnder interferometer (MZI)-assisted microring resonator (MRR I) and a cascaded MRR II, enabling the adaptive generation of response matrices with diverse key metrics using a single device. Our system can be dynamically tuned to capture and encode spectral data with high fidelity and reduced energy consumption, enabling information retrieval comparable to how natural systems discern subtle environmental cues. Crucially, it operates without software intervention, mirroring the autonomy of biological systems while exceeding them in speed and programmability.

By uniting the efficiency of biological encoding with the scalability of photonic technology, our work advances the foundations of spectral analysis and establishes a framework for evolution-aware spectrometer design. Natural principles are not simply imitated but computationally formalized and extended through full-stack innovations in conceptualization, design, implementation, and experiments. Our approach defines a path toward efficient collection and interpretation of spectral data across diverse scientific and industrial applications.

**Conceptualization of bio-inspired generic spectrum encoder**
**Spectrum encoding**

Mathematically, the spectral encoding process can be characterized by a linear mapping $\mathcal{G}_\theta$ ($\mathcal{G}_\theta: L_1(\mathbb{R}_+) \mapsto \mathbb{R}^m$) between a vector space of spectral features $f$ to spectral response $h$, which can be formulated in the form of a Fredholm integral equation of the first kind[28]:

$$h = \mathcal{G}_\theta f = \int_\Omega G(\xi, \theta) f(\xi)\, d\xi \qquad (1)$$

where $\theta := [\theta_1, \theta_2, ..., \theta_q]^T \in \Theta \subseteq \mathbb{R}^q$ with $\Theta$ the corresponding admissible set is the vector of multiple observation channels, such as physical configurations or tunable applied voltages/temperature field on the device; $\xi \in \mathbb{R}_{\geq 0}$ is the wavelength of the admissible spectra. Intuitively, the objective of spectrometer design is to obtain an optimal set of design parameters that enable the observation process to encode as much information as possible. This Expected Information Gains (EIG) can be quantified by the *Kullback-Leibler* divergence[29,30] ($D_{KL}(p\|q) := \mathbb{E}_p[ln(p/q)]$) between the prior density $p(x)$ and the posterior density $p(x|h, \theta)$). Consequently, the quest for optimal design parameters translates into a problem of maximizing Shannon information gain[31] (see Information Theory for Optimal Spectrometer Design in Supplementary Information **S1**):

$$\theta^\dagger := arg\max_{\theta \in \Theta} \text{EIG} = arg\max_{\theta \in \Theta} \mathbb{E}_{h|\theta}[D_{KL}(p(x|h, \theta) \| p(x))] \tag{3}$$

Maximizing EIG with respect to the optimal design parameter $\theta^\dagger$ defines the theoretical optimum. In practice, however, limitations in process technology and restricted tunability in design parameters prevent reaching this optimum, resulting in a trade-off. The whole encoding process is schematically exhibited in Fig. **1a**. We subsequently identify three key metrics for efficient encoding: (1) orthogonality that signifies the insensitivityto noise; (2) completeness that measures the range of spectral sensing, and (3) sparsity that reflects the typicality of the observation matrix and denotes the ability to capture significant features of the spectrum. These three metrics collectively determine the amount of information acquired in measurements and jointly define bounds of the relative error in spectral reconstruction.

To physically realize and experimentally substantiate these metrics, as well as elucidate their interplay, we propose a generic code generator (encoder) with thermally tunable optical characteristics, as briefly schematically illustrated in the upper panel of Fig. **1b**. The device is constructed on a silicon platform and integrated with three TiN microheaters, which allow continuous modulation of the encoder across distinct operational states. Selective activation of individual heaters or their combinations enables adaptive regulation of the above three key metrics. Specifically, orthogonality is governed by the coordinated modulation of Heaters 1 and 2, while completeness is regulated by the collective action of Heaters 1/2 and 3; Sparsity is established through a hybrid hardware–software approach involving Heater 3. These control dependencies are summarized in Fig. **1b**, and the specific tuning mechanisms are elaborated in the '**Demonstration of bio-inspired generic spectrum encoder**' section. Crucially, these three metrics are inherently coupled. Fig. **1a** illustrates this mutual constraint of our system, which serves as a representative model for the trade-offs ubiquitous in spectral encoding systems. We elucidate this theoretical relationship and experimentally validate it in the subsequent sections

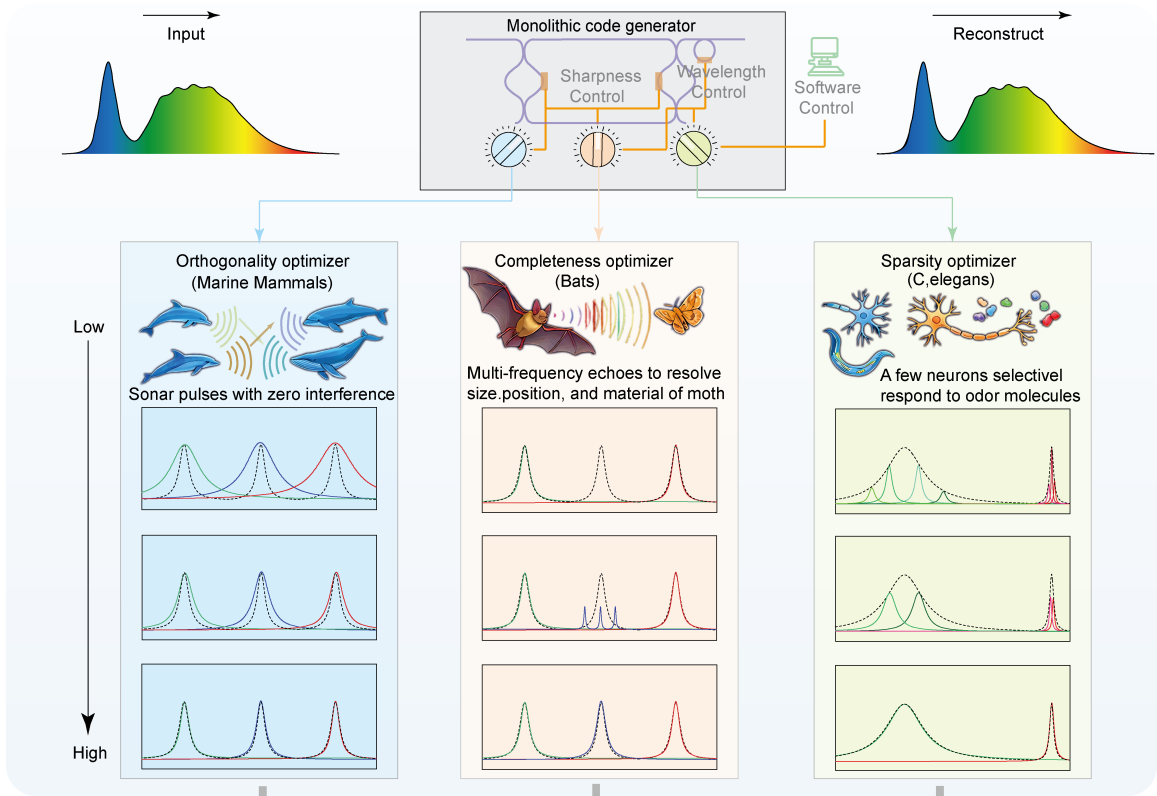

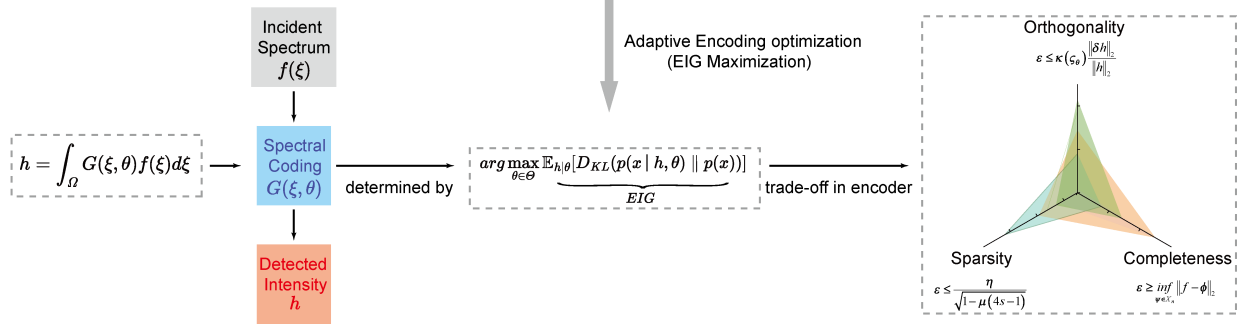

**Fig. 1 | Bio-inspired spectrometer encoding strategy.** (a)Top row, Schematics of the bio-Inspired photonic spectral encoder. Left column, orthogonality. Middle column, completeness. Right column, sparsity. (b)Information-theoretic framework

## Bio-inspired information-theoretic framework of the generic spectrum encoder

### Orthogonality

Orthogonality reflects the ability of the observation matrix to encode significant spectral features, enhances robustness against noise, reduces reliance on prior knowledge, and accelerates iterative convergence. Orthogonality also affects the accuracy of computational reconstruction, determining the upper bound of the error in computational reconstruction[32,33]:

$$\varepsilon \leq \underbrace{\kappa(\mathcal{G}_\theta)}_{\text{Orthogonality}} \frac{||\delta h||_2}{||h||_2} \tag{6}$$

where $\kappa(\mathcal{G}_\theta) = ||\mathcal{G}_\theta|| \cdot ||\mathcal{G}_\theta^{-1}|| = \sigma_{max}/\sigma_{min}$ denotes the condition number[32,33]. A higher condition number indicates poorer orthogonality, resulting from increased overlap between response functions $g_i$ and $g_j$. The left column of Fig. **1b** illustrates the communication patterns of marine mammals, exemplifying orthogonality as manifested in natural biological systems. In optical encoders, orthogonality requires: 1) sharp features with significant perturbation between adjacent wavelengths $max|g_i(\lambda_1) - g_i(\lambda_1 + \Delta\lambda)|$, and 2) minimal periodicity to avoid multicollinearity. The left Fig. **1** demonstrates the direct correlation between orthogonality and reconstruction fidelity, showing progressively better recovery of spectral features relative to the target spectrum (dotted line). Detailed information about orthogonality can be found in Supplementary Information **S2**.

**Completeness**

The middle column in Fig. **1b** depicts bats engaging multiple sensory modalities to perceive their environment comprehensively, reflecting completeness in biological systems. In optical systems, completeness reflects the ability of the observation matrix to accurately represent any spectrum within the operational range. It is quantified by the lower bound of reconstruction error:[34]:

$$\varepsilon \geq \underbrace{\inf_{\psi \in X_n} ||f - \Phi||_2}_{\text{Completeness}} = \inf_{\psi \in X_n} \left\| f - \sum_{k=1}^{m} \alpha_k g_k \right\| \tag{7}$$

where the intensity of $f$ is normalized to make sure $||f||_2 = 1$, and

$$\inf_{\psi \in X_n} ||f - \Phi||_2$$

that denoting relative residual is the metric we used to quantify completeness. In the ideal case, the response matrix $f$ can be completely represented as a linear combination of $\{g_k\}_{k=1}^{m}$ with weight coefficients of $\{\alpha_k\}_{k=1}^{m}$.

An optimal spectral response with adequate completeness should encapsulate sufficient and comprehensive information in response matrix in both row and column direction (each row corresponds to a sampling channel and each column corresponds to a wavelength channel). Thus, any spectrum within the measured spectral range can be fully reconstructed without missing any information. As illustrated in the middle column of Fig. **1b**, the loss of information in the response function at the wavelength of the incident signal leads to incomplete feature extraction, and the spectral peak becomes undetectable in the reconstructed result with the erosion of completeness.

**Sparsity**

The right panel of Fig. **1b** depicts the nervous system of *C. elegans*, where a single neuron is capable of multimodal sensing, efficiently detecting diverse stimuli such as odor molecules, thus reflects its ability to sustain responsiveness while minimizing unnecessary energy expenditure. Sparsity fundamentally reflects the purity of the information content over the collaboration of hardware and software $\tilde{G} = \mathcal{G}_\theta \Psi$.

Sparsity qualifies both the observation matrix $\mathcal{G}_\theta$ provided by the encoder hardware and basis functions $\Psi$ that are generated via software. The sparsity level of $\tilde{G}$ can be measured by the maximum number of mutual coherence $\mu(\tilde{G})$ of $\tilde{G}$. The upper bound of reconstruction error considering sparsity can be estimated as:

$$\varepsilon \leq \frac{\eta}{\sqrt{1-\mu(4s-1)}} \tag{8}$$

where the intensity of $f$ is normalized to make sure $\|f\|_2 = 1$, $\eta$ related to the measurement noise and $s$ denotes the sparse representation of $f$ (detailed in Supplementary Information **S2**).

Considering the optical encoder system, an improvement in sparsity corresponds to a reduction in the inclusion of extraneous elements within the reconstructed spectrum. Consequently, this enhances the purity of the information content within the observation matrix that encapsulates the reconstructed spectrum, as depicted in the right column of Fig. **1b**.

**Theoretical design of bio-inspired generic spectrum encoder**

From a theoretical perspective, orthogonality, completeness, and sparsity jointly determine the range of the error band. Improved performance in these three characteristics leads to higher reconstruction accuracy. To illustrate this relationship and validate the correctness and effectiveness of the theory, we numerically generated matrices with varying properties and reconstructed target spectra from each case, the detailed illustrations of these matrices refer to Supplementary Information **S3**. Furthermore, we analyze error bounds for three attributes based on these matrices, see verifications in Supplementary Information **S4**.

The condition numbers of the numerically generated matrices (models O1–O3, Fig. S2), serving as a metric for orthogonality, are summarized in the left column of Fig. 2a, with values of $1.547 \times 10^{18}$, $2.694 \times 10^{17\,17}$, and $5.573 \times 10^{15}$, respectively. Smoother spectral responses resulted in higher condition numbers and worse orthogonality. This impairment reduced the ability to resolve two close peaks, ultimately causing them to merge into one (Fig. **2a**, right column O1-O3).

The relative residual of numerically generated matrices (models C1–C3, Fig. S2), serving as a metric for completeness, is summarized in the left column of Fig. **2b**. To decouple the effects of individual properties, we hold the orthogonality and sparsity constant for these generated matrices C1–C3. Completeness reflects whether the matrix captures information across the full vector space. As the relative residual increased from 0.034 to 0.432, the weakened left peak became harder to distinguish, along with consistently rising reconstruction error (Fig. **2b**, right column C1-C3).

Regarding sparsity, which indicates both the purity of the hardware-acquired matrix and the effectiveness of basis functions (software), we control the $\mu$ values rose from 0.501 to 0.605, and finally to 0.619 (matrices S1-S3, Fig. **S2**). Corresponding reconstruction errors were 0.034, 0.39, and 0.60 (Fig. 2c, right row). Higher sparsity (i.e., higher $\mu$) reduced accuracy, introduced minor clutter, and increased signal intensity, but ensured no critical signals were omitted. See Supplementary **S2** for sparsity illustrated via basis functions.

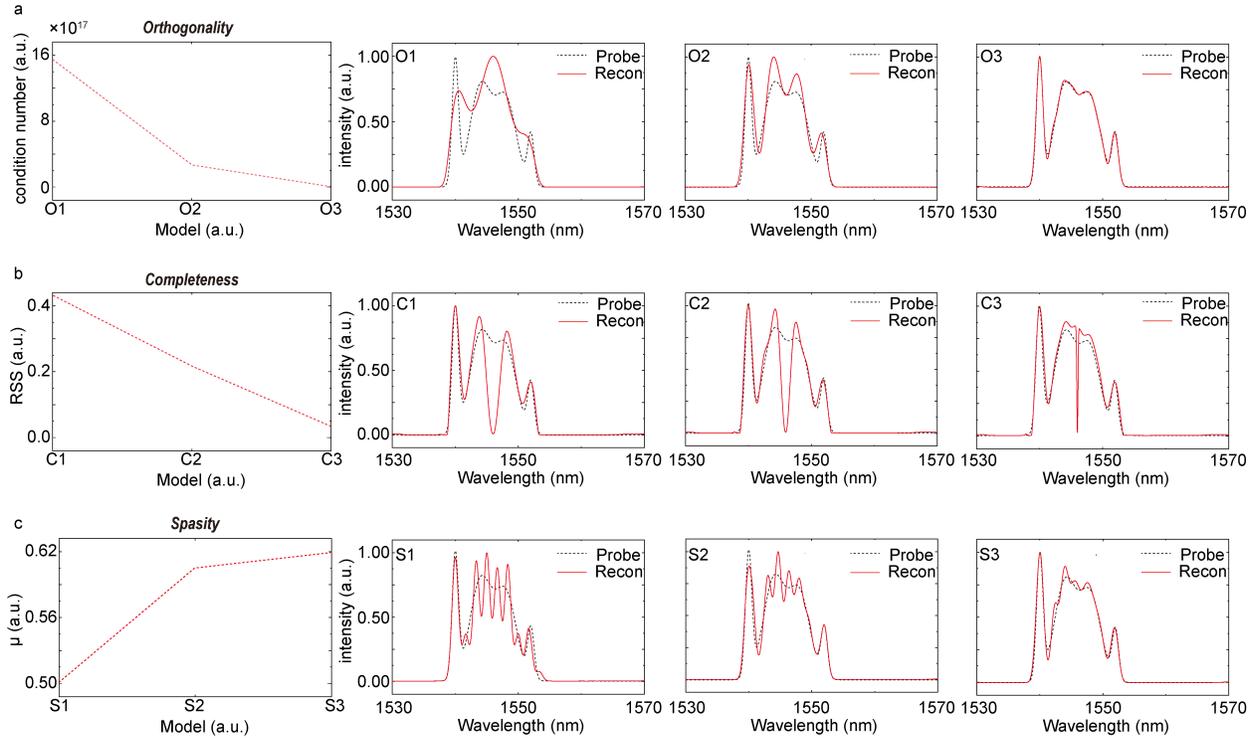

**Fig. 2 | Tunable orthogonality, completeness, and sparsity simulation. a** to **c,** Condition numbers, relative residual and of $\mu(\tilde{G})$ of numerically generated response matrices with improved orthogonality, completeness and sparsity (left columns in each figure), and corresponding reconstruction results (right columns in each figure). The regularization coefficients are fixed as $\alpha_1 = 1\times10^{-10}$ for the $l_1$-regularization term and $\alpha_2 = 1\times10^{-3}$ for the $l_2$-regularization term.

**Demonstration of bio-inspired generic spectrum encoder**

We propose a generic encoder that consists of a high-Q MZI-assisted MRR I and a cascaded MRR II, as exhibited in Fig. **3a**, enabling the adaptive generation of response matrices with diverse key metrics using a single device. Leveraging the Vernier effect in a cascaded MRR II configuration, we achieved a 15-fold extension of the free spectral range (FSR) to approximately 19.2 nm. As illustrated in Fig. **3a**, TiN Heater 1 and Heater 2 located atop the MZI arm of MRR I regulate the bandwidth of the transmission spectrum while Heater 3 in MRR II ring loop tunes the resonance wavelength, thereby configuring each row of the observation matrix. The straight bus waveguides for MRR I are substituted by a pair of equal interferometric arms. Via thermo-optic (TO) effect, the phase difference between the two arms can be continuously tuned to change the effective coupling coefficient $k_{eff}$ for this balance MZI-assisted microring resonator. The variation of the effective coupling coefficient and the corresponding transmissions from the drop port of MRR I (the channel 1 in Fig. **3a**) are simulated and illustrated in Fig. **3b** and **c**, respectively. Therefore, by modulating the external voltages $V_1$, $V_2$, and $V_3$ of three heaters, the transmission spectra of the hardware encoder can be flexibly and adaptively tuned. The orthogonality is governed by the MZI-assisted MRR I, which enables precise adjustment of the spectral response linewidth through the coordinated tuning of Heater 1 and Heater 2; The sparsity is dictated by the cascaded MRR II via tuning Heater 3, which governs information redundancy; while the completeness is governed by the functionality of both the MZI-assisted MRR I and the cascaded MRR II via tuning Heater 1/2 and 3. These functional correspondences are schematically visualized in Fig. **1b**. Details of fabrication and packaging refer to **Methods**. Furthermore, achieving ultra-narrow linewidth spectral response necessitates meticulous loss reduction to enhance the intrinsic Q-factor, which is accomplished via the implementation of a multimode waveguide racetrack configuration and rib waveguides via a shallow-etched strategy (details can be found in Supplementary Information **S5**).

We firstly validate the ability of generating spectral response matrices with diverse degree of orthogonality, completeness, and sparsity for the generic encoder. What is worth noting is that the matrices are reconfigured leveraging the thermo-optic (TO) effect, which enables continuous tuning of their states. For easier clarification, we characterized the encoder under five distinct, predefined configurations of Heater 1 and Heater 2, applied via fixed bias voltages. Condition 1 corresponds to the device's capability to achieve the broadest linewidth (lowest Q-factor), whereas Condition 5 essentially aligns with the device's potential to attain the narrowest bandwidth (highest Q-factor). The power applied on Heater 3 is linearly swept for obtaining an observation matrix where each row corresponds to a discrete sampling channel of $V_3$ and each column represents a wavelength grid. Fig. **3e** exhibits the response matrices obtained under the five conditions with the same $V_3$ configuration. The measured bandwidth, which is the FWHM of resonant peaks in spectral response, is observed to decrease from Conditions 1 to 5. The correlation function $C(\Delta\lambda, N)$ of the spectral response is represented in Fig. **3f** and the estimated resolution can be obtained by the FWHM of $C(\Delta\lambda, N)$. As illustrated in Fig. **S4**, the value and the tendency of measured bandwidth and estimated resolution in the five conditions are consistent.

We subsequently assess the orthogonality, completeness and sparsity of response matrices in these five selected conditions, as exhibited in Fig. **3g**. To evaluate orthogonality, the condition numbers of the physically obtained observation matrices $G_\theta$ from Conditions 1 to 5 are calculated and plotted on the coordinate axis of "orthogonality" in Fig. **3g**. For the evaluation for completeness, a set of representative continuous signals with different center wavelengths, bandwidths and diverse features (see some spectra in Supplementary Information **S7**) is generated by a waveshaper as probe signals. The non-negative least-square method is utilized to calculate relative residual values by equation (7) to qualify the completeness of observation matrices from Conditions 1 to 5. The average calculated relative residual values are presented on the coordinate axis of "completeness" in Fig. **3g**. Because the mutual coherence $\mu$ could reflect the sparsity of the collaboration of hardware and software, $\mu$ of the collaboration of hardware and software $\tilde{G}$ from Conditions 1 to 5 for the assessment of sparsity are calculated and plotted on the coordinate axis of "sparsity" in Fig. **3g.** Here, the basis functions are selected as Gaussian functions. Combining with Fig. **3e, 3g** and Fig. **S4**, it is observed that orthogonality improves gradually as the bandwidth decreases, completeness deteriorates gradually as the bandwidth decreases and sparsity improves gradually as the bandwidth decreases. The experimentally observed matrices generated by a monolithic encoder validate the tunable attributes, including orthogonality, completeness, and sparsity. Still, there exists a certain inherent trade-off between these three characteristics in the generic encoder; for example, optimizing for comprehensive information capture (completeness) often comes at the expense of sparsity and spectral purity.

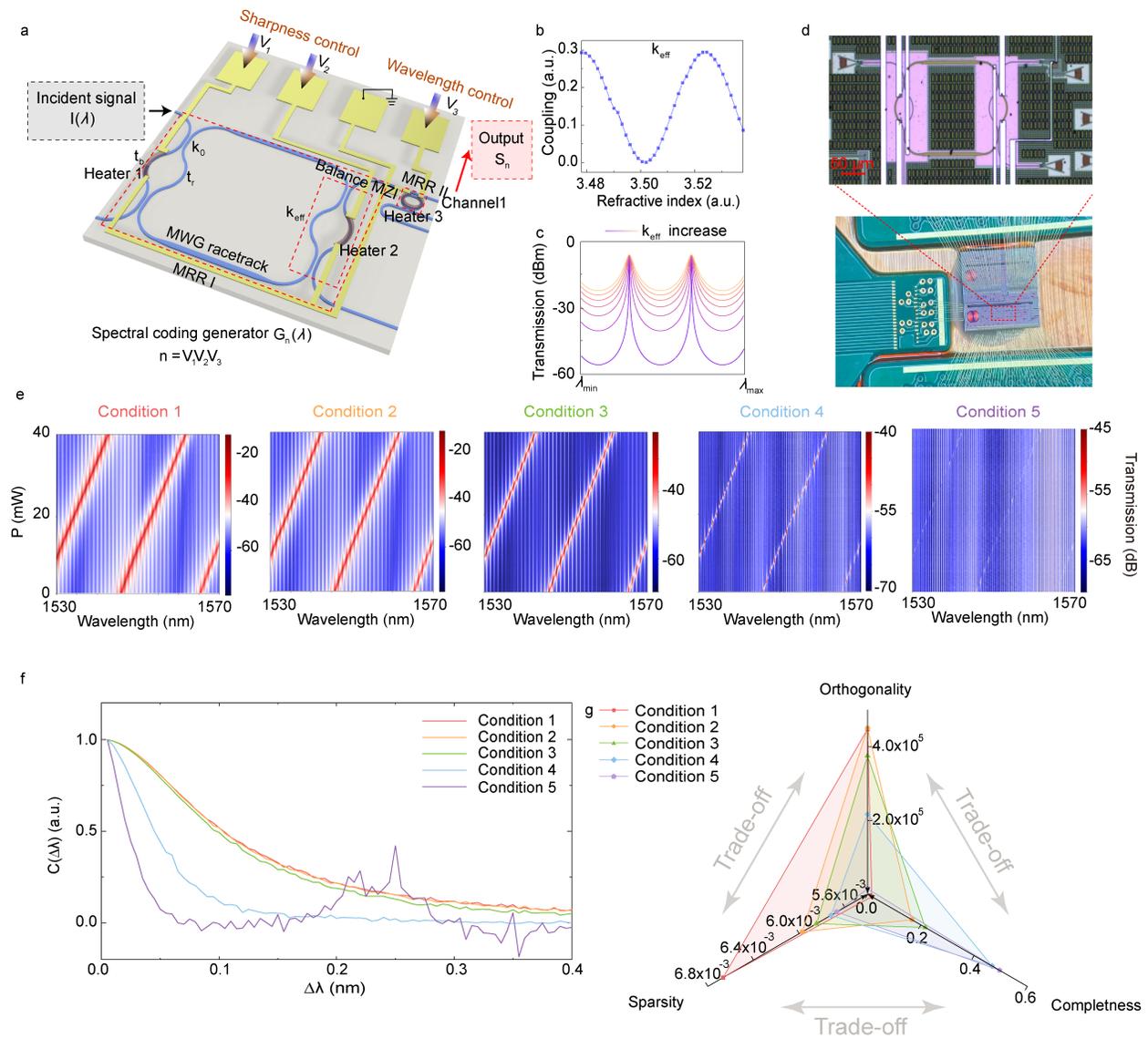

**Fig. 3 | Bio-inspired generic spectrum encoder and the tunable five conditions. a,** Schematic of the encoder. **b,** Simulated effective coupling efficiency as a function of changed refractive indices due to TO effect. **c,** Simulated transmission spectrum from the drop port of MRR with bandwidth tunability. **d,** Microscopy images of the encoder (top) and after the electrical package (bottom). **e,** Response matrices under five conditions. **f,** Correlation functions of response matrices reflect estimated resolution and measured peak bandwidths under five conditions. **g,** Calculated orthogonality, completeness, and sparsity of response matrices under five conditions.

**Reconfigurable properties**

We then validate the synergistic evolution of orthogonality and completeness across distinct condition states. The tendency of discriminative capability, which reflects orthogonality, is experimentally explored using two closely spaced laser signals to probe different matrix conditions. Under optimal orthogonality conditions, an ultra-high spectral resolution of 5 pm was achieved (Supplementary Information **S9**, simulations are also exhibited in Supplementary Information **S8**). Subsequently, a smooth continuous signal that generated by an ASE laser source, spectrally shaped by a bandpass filter, is utilized to probe the tendency of completeness under different conditions. Simulation and experimentally reconstructed results are provided in Supplementary Information **S9**, Fig. **S7b**. Besides, the experimental reconstruction results for single-peak signals that validated the completeness of discrete signals is also verified experimentally in Supplementary Information **S11**, which also shows the intertwined instinct of completeness and sparsity. The above results reveal an intrinsic trade-off: as the spectral linewidth narrows, the resolution of dual-peak signals improves, whereas the fidelity of smooth signal reconstruction degrades, confirming that enhancements in orthogonality are inversely correlated with completeness. For sparsity, we explore the impact of the basis function determined by the software on sparsity, and to investigate its relationship with reconstruction accuracy. Relative results are provided in Supplementary Information **S9**.

**Validation and Testing – reconfigurable attribute and versatile response functions library**

For coping with incident signals with varied and complicated optical features, we employ programmable random sequences ($V_1/V_2$) in our generic encoder to generate a versatile library of response functions with different orthogonality, completeness and sparsity. Under the programmable random sequences, the response matrix is measured from the linearly swept heating power of the Heater 3. Fig. **4a** exhibits the measured response matrix under 200 heating channels with a wavelength grid of 5 pm. Fig. **4b** exhibits the ultra-low level of cross correlation for measured response matrix. As detailed in Supplementary Information **S12**, the observation matrix further exhibits favorable properties in terms of correlation (the calculated average auto-correlation function reveals the low periodicity level of 0.22), condition number, etc. Leveraging the enhanced properties of the random-coded matrix, the measurable bandwidth is extended to more than twice the original FSR. To explore richness of the matrix, we further quantify the orthogonality, completeness, and sparsity metrics for distinct subsets within the matrix, each comprising five randomly selected response functions. The distributions of these computed metrics, presented in Fig. **4c**, exhibit significant diversity. This heterogeneity of attributes demonstrates that this randomly encoded matrix encompasses a diverse repertoire of functions with distinct properties.

We then experimentally validate the performance of the proposed spectrometer with respect to a range of different characterization challenges. We firstly test the resolution limit of our device through reconstructing the simultaneously launching two laser signals, which are separated by 6 pm, 20 pm, 10 pm and 0.84 nm, as illustrated in Fig. **4d**, demonstrating the ultra-high resolution of 6 pm in our device. For continuous signal reconstruction, compared to the references calibrated from the optical spectrum analyzer which are marked as the black dashed line, various continuous signals with diverse complex spectral features are well reconstructed, as shown in Fig. **4e** (details refer to **Methods**). The calculated average relative errors ($\varepsilon = \frac{\|I-I^\dagger\|_2}{I}$) and coefficient of determination ($r^2 = 1 - \frac{\Sigma_1^n(I-I^\dagger)^2}{\Sigma_1^n(I)^2}$) is labeled in Fig. **4e**, indicating high fidelity.

Consequently, due to the inclusion of a diverse range of functions with varying attributes (orthogonality, completeness and sparsity) within this library, our device can be perfectly adapted to input signals with different optical features without prior knowledge, overcoming the longstanding resolution-reconstruction trade-off in conventional spectrometers and achieving superior fidelity across both continuous and discrete spectral regimes with an ultra-high resolution of 6 pm and a wide operational bandwidth of 30 nm. Furthermore, the construction of this library of response functions offers the potential for more flexible and adaptive applications in spectral testing. By modulating the state of the encoder correspondingly, based on the known general characteristics of the input signal (e.g., whether it is sparse or continuous), the corresponding matrix can be generated, enabling efficient and perfect reconstruction of the spectrum.

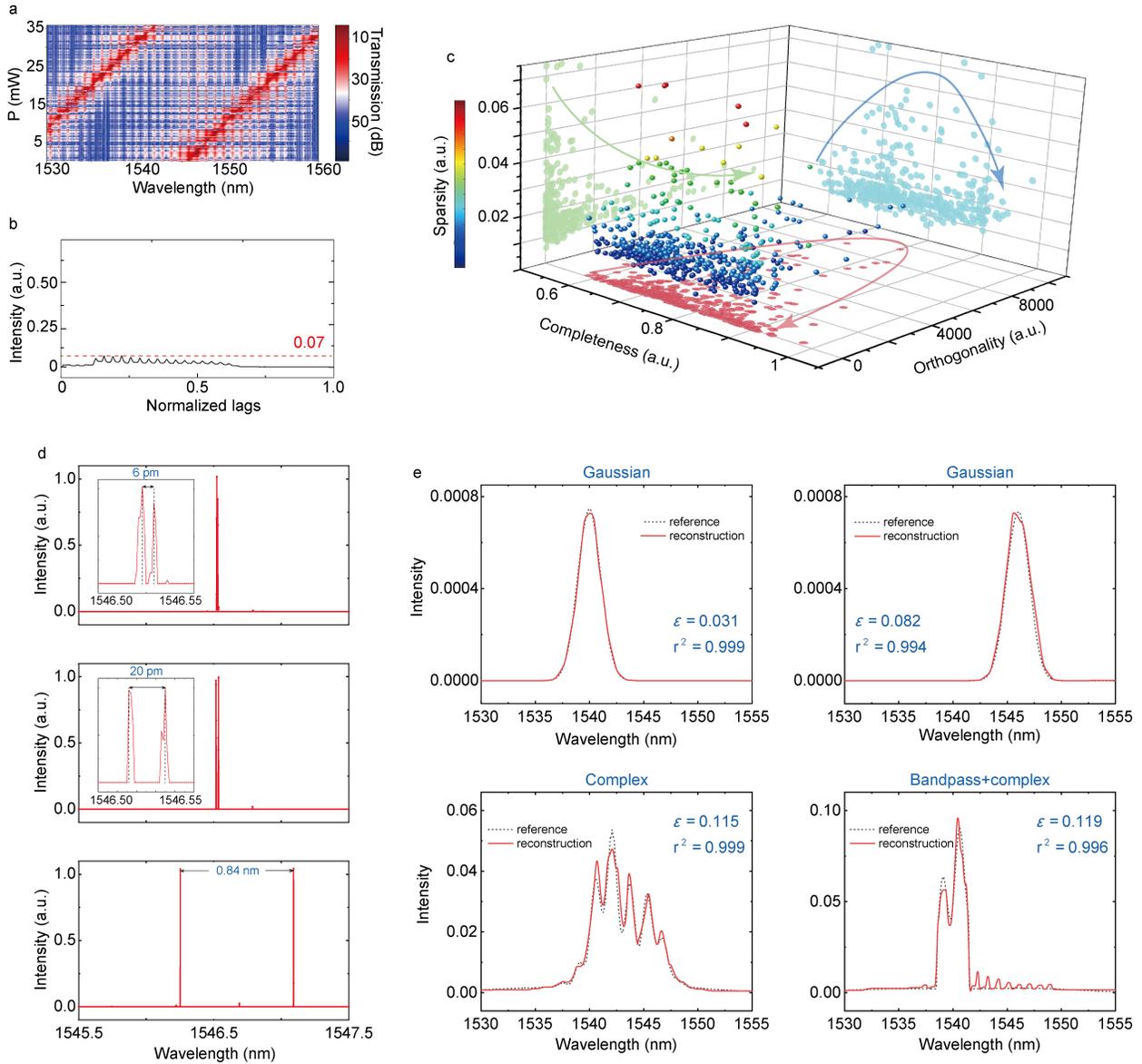

**Fig.4 | Matrix and reconstruction results of bio-inspired generic spectrum encoder. a,** Response matrix under linearly swept heating power from 0 to about 35 mW with 200 heating channels from the Heater 3. **b,** The average cross correlation function of response matrix. **c,** Distributions of three attributes of the subset in the matrix. Demonstrations of the spectrometer's reconstruction capabilities with respect to **d,** double-peak resolving that the minimum differentiable peak separation of 6 pm and **e,** continuous signal with Gaussian, complex and combination of bandpass and complex spectra features.

**Discussion and conclusion**

Optical spectroscopy has long been constrained by the lack of a unified encoding framework addressing the inherent complexities of diverse spectral datasets. In this work, inspired by the biological encoders, we introduce a generic spectral encoder that bridges this gap, transforming a conventional information-theoretic approach into a dynamic, scalable platform for spectrometer design. By optimizing the three critical physical attributes of orthogonality, completeness, and sparsity within a single, reconfigurable architecture, we address long-standing constraints in hardware flexibility. Through a rigorous survey of the state-of-the-art, we present the first dynamically reconfigurable spectrum encoder capable of modulating observation matrices via external voltage control. This architecture pioneers adaptive measurement directly at the hardware level. Details of performance benchmarking can be found in Supplementary Information **S13**. This fundamental innovation enables the adaptive tuning of spectral properties to accommodate a broad range of signals, from smooth, broadband spectra to discrete peaks with complex features. Crucially, this adaptive capability offers unprecedented versatility in spectral analysis, ensuring optimal performance across diverse and challenging datasets.

This work marks a significant step forward in the unification of theoretical precision with experimental adaptability, laying the foundation for spectrometers that can dynamically tailor their encoding strategies in real-time, much like biological systems. This flexibility opens the door to a new era of intelligent, adaptive spectroscopic sensing that could revolutionize applications in fields as varied as portable biomedical diagnostics, environmental monitoring, and space-based spectral imaging. Beyond its technical achievements, the implications of this work extend to the democratization of high-precision spectroscopy, making advanced sensing capabilities more accessible and scalable for a wide range of real-world applications. The ability to reconfigure spectrometers on-the-fly represents a leap towards ubiquitous, intelligent sensing systems capable of tackling the increasingly complex demands of modern science and technology.

## Methods

### Device fabrication.

The encoder is fabricated in Chongqing United Microelectronics Center (CUMEC) using their CSiP180A technology with 248nm lithography. The multi-project wafer (MPW) starts with silicon-on-insulator (SOI) 200mm wafers using 220nm top silicon and 2μm buried oxide (BOX). After 1 mm oxide cladding 120 TiN metal heater and aluminum metal interconnection with a thickness of 1 mm are formed, subsequently.

### Calibration.

A pre-calibration process is demonstrated as follows. A tunable continuous wave laser (Santec TSL 770) and a power monitor (Santec MPM 210) are utilized for sampling and data collection. Broadband grating couplers (GC) with less than 5 dB insertion loss are applied for fiber-chip coupling. A source meter (Keithley 2400) is used to offer an external driving power source. A maximum external power $P_{max}$ of 60 mW is linearly swept and applied to Heater 3. The other two source meter is used to heat the Heater 1 and 2 to set the matrix in different conditions. For eliminating current fluctuations due to unstable contact, we utilize wire bonding and an electrical package for our fabricated chip. The current value can be stabilized at 1×10⁻⁴ mA after electrical packaging. The dimensions of the response function T are determined by the wavelength point number Mw = BW/δλ, and Np = Pmax/δP mutually, where Np is the sampling channels of 300; BW is the spectrometer measurable bandwidth (1500 nm to 1600 nm); δλ is the wavelength grid of 10 pm.

For dual-peak reconstruction, we use two CW laser sources (Santec TSL 770) to generate discrete narrow linewidth signals with different wavelength separation of 5 pm, 10 pm and 20 pm, which are combined by a 3-dB coupler. For continuous signal reconstruction, a broadband amplified spontaneous emission (ASE) source is utilized to supply incident continuous light centered at 1550 nm. A commercial optical filter is connected to generate the continuous band-pass signals. An Erbium-Doped Fiber Amplifier (EDFA) is connected to amplify signal light. For continuous signal reconstruction, we use a commercial optical spectrum analyzer (OSA, Yokogawa AQ6370C) to collect the optical response of the wave-shaper (Finisar Wave-shaper 1000s) that is coded with the required signals in advance for calibration and reference. Continuous signals with Gaussian, complex and superposition of bandpass filter and complex spectral features are generated.

The spectrum reconstructions are performed by running the CVX optimization algorithm on MATLAB by the "Mosek" solver, based on an AMD Ryzen 7 3700X CPU and NVIDIA GeForce GTX 1650 with 32 GB memory. The reconstruction of continuous signals is achieved in approximately 3 seconds, whereas the reconstruction of discrete sparse signals requires approximately 2 seconds. ANSYS Lumerical FDTD and INTERCONNECT are used to perform the optical transmission simulations. For the spectrum measurement process, we sample every 0.5 s with an integration time of 0.1 s. Since the rise time and the fall time are less than 10 μs, sampling starts at 0.3 s after sending commands to the source meter to regulate the external biases, in order to ensure a stable temperature of the silicon waveguide.

**Reconstruction algorithms.**

The reconstructed signal $I^\dagger$ can be generated by solving the under-determined linear least-squares:

$$I^\dagger = argmin_{I \in \mathbb{R}^+} \|TI - S\|_2^2 \tag{14}$$

Where $\|\cdot\|_2$ represents the $l_2$-norm. It is noteworthy to mention that the response matrix obtained in real-world circumstances frequently indicates a certain level of ill-conditioning, particularly in regard to the interpretation of different incident signals in the presence of measurement noise. In order to overcome this overfitting obstacle, the regularization coefficient is introduced as:

$$I^\dagger = argmin_{I \in \mathbb{R}^+}(\|T \times I - S\|_2^2 + \alpha_1\|I\|_1 + \alpha_2\|I\|_2 + \alpha_3\|DI\|_2), 0 \leq I \leq 1 \tag{15}$$

where $\alpha_1$ is the weight regularization coefficient for the l₁-norm of the input matrix *I*, which is vital in the regression of $I^\dagger$ into discrete untrivial solutions; a₂ is the weight regularization coefficient for the l₂-norm of *I*, which strengthen the robustness against measurement noise and decrease complexity; D refers to the first derivative operator; a₃ is the weight regularization coefficient for the l₂-norm of the first derivation of *I*, which is critical for optimizing continuity and smoothness of continuous broad-band signals CVX optimization algorithm implemented in MATLAB environment is utilized to solve this convex optimization.

**Acknowledgments**

This work was supported by the National Research and Development Program of China (2023YFB2804702); National Natural Science Foundation of China (NSFC) (62550072, 62341508); National Science Foundation of China (U24A2057); Shanghai Science and Technology Innovation Action Plan (25LN3201000 and 25JD1405500); Scientific Research Innovation Capability Support Project for Young Faculty (ZYGXQNJSKYCXNLZCXM-M18); Shanghai Municipal Science and Technology Major Project. We also thank the Center for Advanced Electronic Materials and Devices (AEMD) of Shanghai Jiao Tong University (SJTU) and Chongqing United Microelectronics Center (CUMEC) for fabrication support.

**Author contributions**

X. G. initiated the project. Y. Z., X. L., Y. C. and C.X. performed the calculation and simulation. X.G. and Y. Z. designed the experiments. Y. Z. fabricated samples. Y. Z. and C.X. carried out the measurements. X.G., Q.Z., Y. Z., X. L., Y. C., H. C., Z. Y., Y.S., Z.S., and T. H. analyzed the results and wrote the manuscript. X. G., H. C., Z. Y., Y.S., Z.S., and T. H. supervised the project.

**Competing interests**

The authors declare no competing interests.

**Data availability**

The data is available via Zenodo at https://doi.org/10.5281/zenodo.17907415.[35]

**Code availability**

The software code is available via Zenodo at https://doi.org/10.5281/zenodo.17907415.[35]

**References**
1   Bialek, W. *Biophysics: searching for principles*.   (Princeton University Press, 2012).
2   Menze, S., Zitterbart, D. P., van Opzeeland, I. & Boebel, O. J. R. S. O. S. The influence of sea ice, wind speed and marine mammals on Southern Ocean ambient sound.   **4**, 160370 (2017).
3   Stein, B. E. & Stanford, T. R. J. N. r. n. Multisensory integration: current issues from the perspective of the single neuron.   **9**, 255-266 (2008).
4   Liu, Q., Kidd, P. B., Dobosiewicz, M. & Bargmann, C. I. J. C. C. elegans AWA olfactory neurons fire calcium-mediated all-or-none action potentials.   **175**, 57-70. e17 (2018).

5       Sung, S. H. *et al.* Bio-plausible memristive neural components towards hardware implementation of brain-like intelligence.   **62**, 251-270 (2023).
6       Wang, B. *et al.* Biologically inspired heterogeneous learning for accurate, efficient and low-latency neural network.   **12**, nwae301 (2025).
7       Bao, J. & Bawendi, M. G. J. N. A colloidal quantum dot spectrometer.   **523**, 67-70 (2015).
8       Yang, Z. *et al.* Single-nanowire spectrometers.   **365**, 1017-1020 (2019).
9       Yoon, H. H. *et al.* Miniaturized spectrometers with a tunable van der Waals junction.   **378**, 296-299 (2022).
10      Tang, F. *et al.* Metasurface spectrometers beyond resolution-sensitivity constraints.   **10**, eadr7155 (2024).
11      Zhang, G. *et al.* Stress-engineered ultra-broadband spectrometers.   **11**, eadu4225 (2025).
12      Yuan, S., Naveh, D., Watanabe, K., Taniguchi, T. & Xia, F. J. N. P. A wavelength-scale black phosphorus spectrometer.   **15**, 601-607 (2021).
13      Fan, Y. *et al.* Dispersion-assisted high-dimensional photodetector.   **630**, 77-83 (2024).
14      Cai, G. *et al.* Compact angle-resolved metasurface spectrometer.   **23**, 71-78 (2024).
15      Redding, B., Liew, S. F., Sarma, R. & Cao, H. J. N. P. Compact spectrometer based on a disordered photonic chip.   **7**, 746-751 (2013).
16      Redding, B., Fatt Liew, S., Bromberg, Y., Sarma, R. & Cao, H. J. O. Evanescently coupled multimode spiral spectrometer.   **3**, 956-962 (2016).
17      Hadibrata, W. *et al.* Compact, high‑resolution inverse‑designed on‑chip spectrometer based on tailored disorder modes.   **15**, 2000556 (2021).
18      Xu, H., Qin, Y., Hu, G., Tsang, H. K. J. L. S. & Applications. Breaking the resolution-bandwidth limit of chip-scale spectrometry by harnessing a dispersion-engineered photonic molecule.   **12**, 64 (2023).
19      Xu, H., Qin, Y., Hu, G. & Tsang, H. K. J. O. Cavity-enhanced scalable integrated temporal random-speckle spectrometry.   **10**, 1177-1188 (2023).
20      Yao, C. *et al.* Broadband picometer-scale resolution on-chip spectrometer with reconfigurable photonics.   **12**, 156 (2023).
21      Yao, C. *et al.* Integrated reconstructive spectrometer with programmable photonic circuits.   **14**, 6376 (2023).
22      Zhao, Y. *et al.* Miniaturized computational spectrometer based<? TeX\break?> on two-photon absorption.   **11**, 399-402 (2024).
23      Zhang, Y. *et al.* Miniaturized disordered photonic molecule spectrometer.   **14**, 144 (2025).
24      Zhang, Y. *et al.* Miniaturized chaos-assisted Spectrometer.   **14**, 331 (2025).
25      Hao, L. *et al.* Self‑Adaptive Miniaturized Spectrometer Leveraging Wavelength‑Tunable Organic Photodetectors for High‑Resolution Spectral Sensing. 11847 (2025).
26      Tang, H. *et al.* An adaptive moiré sensor for spectro-polarimetric hyperimaging. 1-8 (2025).
27      Tang, H. *et al.* An adaptive moiré sensor for spectro-polarimetric hyperimaging. *Nature Photonics* **19**, 463-470 (2025). https://doi.org:10.1038/s41566-025-01650-z
28      Fredholm, I. Sur une classe d'équations fonctionnelles.   (1903).
29      Kullback, S. & Leibler, R. A. On information and sufficiency. *The annals of mathematical statistics* **22**, 79-86 (1951).
30      Kullback, S. *Information theory and statistics*.   (Courier Corporation, 1997).
31      Chaloner, K. & Verdinelli, I. Bayesian experimental design: A review. *Statistical science*, 273-304 (1995).
32      Belsley, D. A., Kuh, E. & Welsch, R. E. *Regression diagnostics: Identifying influential*

|     | *data and sources of collinearity*. (John Wiley & Sons, 2005). |
| --- | --- |
| 33 | Golub, G. H. & Van Loan, C. F. *Matrix computations*. (JHU press, 2013). |
| 34 | Kress, R. *Linear integral equations*. Vol. 82 (Springer, 1999). |
| 35 | Zhang, Y. (2025). Bio-inspired Photonic Spectral Encoders. Zenodo. https://doi.org/10.5281/zenodo.17907415 |